\documentclass[12pt]{article}
\usepackage{amsmath}
\usepackage{amssymb}
\usepackage{citesort}
\usepackage{graphicx}
\usepackage[nomarkers]{endfloat}
\usepackage{pstricks}
\numberwithin{equation}{section}
\begin{document}

\setcounter{page}{0}
\thispagestyle{empty}

\begin{flushright}
{\small BARI-TH 427/01}
\end{flushright}

\vspace*{2.5cm}

\begin{center}
{\large\bf Dynamics of Walls: Fermion Scattering
\\[0.4cm] in Magnetic Field  }
\end{center}

\vspace*{2cm}

\renewcommand{\thefootnote}{\fnsymbol{footnote}}

\begin{center}
L. Campanelli$^{1,2,}$\protect\footnote{Electronic address: {\tt
leonardo.campanelli@ba.infn.it}},
P. Cea$^{1,2,}$\protect\footnote{Electronic address: {\tt
Cea@ba.infn.it}},
G.~L. Fogli$^{1,2,}$\protect\footnote{Electronic address: {\tt
Fogli@ba.infn.it}} and
L. Tedesco$^{1,2,}$\protect\footnote{Electronic address: {\tt
luigi.tedesco@ba.infn.it}} \\[0.5cm] $^1${\em Dipartimento di Fisica,
Universit\`a di Bari, I-70126 Bari, Italy}\\[0.3cm] $^2${\em INFN
- Sezione di Bari, I-70126 Bari, Italy}
\end{center}

\vspace*{0.5cm}

\begin{center}
{%\large
November, 2001 }
\end{center}

\vspace*{1.0cm}

\renewcommand{\abstractname}{\normalsize Abstract}
\begin{abstract}
We investigate the scattering of fermions off walls in the  presence of
a magnetic field. We consider both the bubble wall and the kink
domain wall. By solving the Dirac equation for fermions in
the presence of a domain wall in an external magnetic field, we
investigate the dependence on the magnetic field
of  the transmission and reflection
coefficients. In the case of
kink domain wall, we also consider the solutions localized on the
wall. The possibile role of the ferromagnetic domain walls in the
dynamics of the early Universe is also discussed.
\end{abstract}
\vspace*{0.5cm}
\begin{flushleft}
PACS number(s): 11.15.Ha
\end{flushleft}
\vfill
\newpage
\renewcommand{\thesection}{\normalsize{\Roman{section}.}}
%\begin{center}
\section{\normalsize{INTRODUCTION}}
%\label{INTRODUCTION}
%\end{center}
\renewcommand{\thesection}{\arabic{section}}
Recently, a considerable amount of renewed interest has  emerged in
the physics of topological defects produced during cosmological
phase transitions~\cite{Vilenkin:1994paper}.
%
%\footnote{For a review, see Ref.~\cite{Vilenkin:1994paper}.}
%
It is known since long time that, even in a perfectly homogeneous
continuous phase transition, defects will form if the transition
proceeds sufficiently faster than the relaxation time of the order
parameter~\cite{Kibble:1976sj,Kibble:1980mv,Zurek:1985qw,Zurek:1996sj}.
In such a non-equilibrium transition, the low temperature phase
starts to form, due to quantum fluctuations, simultaneously and
independently in many parts of the system. Subsequently, these
regions grow together to form the broken-symmetry phase. When
different causally disconnected regions meet, the order parameter
does not generally match and a domain structure is formed. 
\\
In the standard electroweak phase transition the neutral Higgs field is
the order parameter which is expected to undergo a continuum phase
transition. Actually, if we compare the lower bound
recently established by the ALEPH
Collaboration~\cite{Barate:2000zr},
\begin{equation}
\label{bound}
M_H \; > \; 107.7 \; \text{GeV} \; \;  \text{at} \;  \; 95 \;  \% \;  \; 
\text{C.L.} \; ,
\end{equation}
with the results of non-perturbative lattice
simulations~\cite{Fodor:1999at}, we are induced to safely exclude a first
order electroweak phase transition~\cite{Kajantie:1996mn}.
However, a first order phase transition can be nevertheless
obtained with an extension of the Higgs sector of the Standard
Model, or in the Minimal Supersymmetric Standard Model.
If we assume that this is the case, it can be also 
conjectured that the  observed 
baryon asymmetry
may be generated at the primordial electroweak phase
transition~\cite{Trodden:1998ym,Riotto:1998bt,Riotto:1999yt}. \\
In the case in which the phase transition is induced 
by the  Higgs sector of the Standard Model, the defects are
domain walls across which the field flips from one minimum to the
other. The defect density is then related to the domain size and the
dynamics of the domain walls is governed by the surface tension
$\sigma$. The existence of the domain walls, however, is still
questionable:
it was pointed out by Zel'dovich,
Kobazarev and Okun~\cite{Zeldovich:1974uw} that the gravitational
effects of just one such wall stretched across the universe would
introduce a large anisotropy into the relic blackbody radiation.
For this reason the existence of such walls was excluded. 
Quite recently, however, it has been suggested
\cite{Cea:1998ep,Cea:1999jz} 
that the effective surface tension of the domain walls can be made
vanishingly small due to a peculiar magnetic condensation induced
by fermion zero modes localized on the wall. As a consequence, the
domain wall acquires a non zero magnetic field perpendicular to
the wall, and it becomes almost invisible as far as the
gravitational effects are concerned.  \\
In a similar way, even for the bubble walls it has been
suggested~\cite{Vachaspati:1991nm} that strong magnetic fields may
be produced as a consequence of non vanishing spatial gradients of
the classical value of the Higgs field. 
Thus, we are led to suppose that in general the magnetic field vanishes in the
regions where the scalar condensate is constant: it can be different 
from zero only in the regions
where the scalar condensate varies, i.e. in a region of the order
of the wall thickness.
\\
It is worthwhile to stress that in the realistic case
where the domain wall interacts with the plasma, the magnetic
field penetrates into the plasma over a distance of the order of
the penetration length, which at the epoch of the electroweak phase
transition is about an order of magnitude greater than the wall
thickness. This means that fermions which scatter on the wall feel
an almost constant magnetic field over a spatial region much
greater than the wall thickness. So that we can assume that the
magnetic field is constant. \\
The aim of this paper is to study the scattering of fermions off
the walls; in particular, within the so-called {\it {defect mediated
electroweak baryogenesis}}. 
Since the dynamical generation of the baryon asymmetry at the
electroweak transition is related to the transmission and
reflection coefficients, it is interesting to see how the
magnetic field localized at the wall modifies these coefficients.
In a previous paper~\cite{Cea:2000wz} we considered Dirac fermions
with  momentum asymptotically  perpendicular to the wall
surface; this corresponds to neglect the motion parallel to the
wall surface. In the present paper we include the effects due to
the motion of fermions parallel to the wall. In particular, we
will see that, in the scattering of fermions off domain walls in
the presence of a constant magnetic field, there are localized states
corresponding to fermions which asymptotically have positive
energy. These localized states are a peculiar
characteristic  of domain walls and are expected to play an important role in
the dynamics of the walls.
\\
The plan of the paper is as follow. In Section~2 we discuss  the
Dirac equation in the presence of a planar domain wall with a constant
magnetic field perpendicular to the wall. We evaluate the
reflection and transmission coefficients and compare them with the
known results without the magnetic field. Section~3 is devoted to
the study of the localized states on the planar domain wall. Finally,
some concluding remarks are presented in Section~4. For
completeness, in the Appendix we discuss the case of bubble walls
in the presence of magnetic field perpendicular to the wall.
\\
%
%
%#######################################################################
%
\renewcommand{\thesection}{\normalsize{\Roman{section}.}}
\section{\normalsize{KINK DOMAIN WALL}}
\renewcommand{\thesection}{\arabic{section}}
In this Section we consider domain walls which are thought to be
formed in a continuous phase transition by the Kibble
mechanism~\cite{Kibble:1976sj,Kibble:1980mv,Zurek:1985qw,Zurek:1996sj}.
When the scalar field develops a non vanishing vacuum expectation
value $\langle \phi \rangle = \pm v$, one may assume
that there are regions with $\langle \phi \rangle = + v$ and
$\langle \phi \rangle = - v$. It is easy to see that the classical
equation of motion of the scalar field admits the solution
describing the transition layer between two adjacent regions with
different values of $\langle \phi \rangle$:
\begin{equation}
\label{2.1}
\varphi(z)= v \, \tanh \left(\frac {z} {\Delta} \right),
\end{equation}
where $\Delta$ is the domain wall thickness, which in the following we
will set to 1.
We are interested in the scattering of Dirac fermions off planar
domain wall in the presence of the electromagnetic field
$A_{\mu}$. Thus, assuming that fermions are coupled to the scalar
field through a Yukawa interaction with coupling $g_Y$, 
the Dirac equation reads:
\begin{equation}
\label{eq2.2}
(i \, \gamma^{\mu} \partial_{\mu} - g_Y \, \varphi - e \, \gamma^{\mu} A_{\mu})
\, \Psi(x,y,z,t) =0 \, ,
\end{equation}
where $e$ is  the electric charge. Putting
\begin{equation}
\label{eq2.3} \xi=g_Y \, v \; \; \; ,\; \; \; \;  g(z) = \tanh \, z\, ,
\end{equation}
where $\xi$ is the fermion mass in the broken phase, Eq.
(\ref{eq2.2}) becomes
\begin{equation}
\label{eq2.4}
(i \, \gamma^{\mu} \partial_{\mu} - \xi \, g(z) - e \, \gamma^{\mu} \, A_{\mu})
\, \Psi(x,y,z,t) = 0 \, .
\end{equation}
In order to solve Eq. (\ref{eq2.4}), we write
\begin{equation}
\label{eq2.5}
\Psi(x,y,z,t) = (i \, \gamma^{\mu} \partial_{\mu} + \xi \, g(z) - e \,
\gamma^{\mu} \, A_{\mu}) \, \Phi(x,y,z,t) \, ,
\end{equation}
and insert  Eq.~(\ref{eq2.5}) into Eq.~(\ref{eq2.4}), using the
standard representation~\cite{Bjorken:1962book} for the Dirac
matrices. We obtain:
\begin{eqnarray}
\label{eq2.6}  & & [ \phantom
[- \partial^2 +
i \, \xi \gamma^3 \partial_z \, g(z) - 2 i \, e
\,A^{\mu}  \partial_{\mu} - i \, e
\partial^{\mu} \, A_{\mu}- \frac {e} {2} \,
\sigma^{\mu \nu}
\, F_{\mu \nu} +
\nonumber
\\
& &
- 
\xi^2 \, g^2(z) + e^2 \,
A_{\mu}
A^{\mu} ] \, \Phi(x,y,z,t) = 0.
\end{eqnarray}
Now, writing
\begin{equation}
\label{eq2.7} \Phi(x,y,z,t)=\chi(x,y,z)\, e^{- i\, E_r \, t}  \; ,
\end{equation}
where
\begin{equation}
\label{eq2.8}
E_r =
         \begin{cases}
            + E \; \; \; \; \text{for}  \; \; \; \;  r=1 \; \; \; \;
               \text{(positive energy solutions)} \\
            - E \; \; \; \;  \text{for} \; \; \; \;  r=2 \; \; \; \;
                \text{(negative energy solutions)} \; \; ,
         \end{cases}
\end{equation}
we get:
\begin{eqnarray}
\label{eq2.9} & & \left[ \phantom{\frac {1} {2}}
 E^2 +
\partial^2_k  - 2 \, i \, e \, A^k
\partial_k + e^2 A_k A^k   - \xi^2 g^2(z) \, +
\right. \nonumber
\\
& &  + \left. i \, \xi \gamma^3 \partial_z g(z) - \frac {e} {2} \,
\sigma^{\mu \nu} F_{\mu \nu} \right]\chi(x,y,z) \, e^{- i \, E_r
\, t} =0 \; .
\end{eqnarray}
According to our previous discussion, we can safely assume that
$A_{\mu} $ corresponds to a constant magnetic field directed along
$z$ with strength $F_{2 1}=B$. We can then choose the gauge
such that:
\begin{equation}
\label{eq2.10} A_{\mu} = (0,0,-Bx,0) \; .
\end{equation}
Setting now:
\begin{equation}
\label{eq2.11} \chi(x,y,z)=f(x,y) \, \omega(z) \; ,
\end{equation}
we see that  Eq.~(\ref{eq2.9}) gives rise to two independent
equations:
\begin{equation}
\label{eq2.12}
[ \, \partial_x^2 + \partial_y^2
- 2 \, i \, e \, A^2 \, \partial_y - e^2 \, (A_2)^2
+ E_{\perp} \, ] \, f(x,y)= 0
\end{equation}
and
\begin{equation}
\label{eq2.13} [\, \partial^2_z + i \, \xi \, \gamma^3 \,
\partial_z \, g(z) - \xi^2 \, g^2(z) + {E'}^2 + i \, e \, B \,
\gamma^1 \, \gamma^2 \, ] \, \omega(z) =0 \, ,
\end{equation}
with
\begin{equation}
\label{eq2.14}
E^2={E'}^2\, + E_{\perp} \; .
\end{equation}
Let us consider first Eq.~(\ref{eq2.12}). With the substitution 
\begin{equation}
\label{eq2.15}
f(x,y) = e^{i \, p_y \, y} \, h(x)\, ,
\end{equation}
 Eq.~(\ref{eq2.12}) becomes:
\begin{equation}
\label{eq2.16}
\left[ - \frac {d^2} {d x^2} + p^2_y - 2 \, e \, B \,  x \, p_y \,+
e^2\, B^2 \, x^2 \right] h(x) = E_{\perp} \, h(x) \, .
\end{equation}
With the change of variable
\begin{equation}
\label{eq2.17}
\zeta= \sqrt{e\, B} \left( x- \frac {p_y} {eB}\right),
\end{equation}
it is straightforward to show that the solutions of the
Eq.~(\ref{eq2.16}) can be written in the form
\begin{equation}
\label{eq2.18}
h_n(\zeta) = A_n \, e^{- \frac {1} {2} \zeta^2 } \, H_n(\zeta)\, ,
\end{equation}
where $H_n(\zeta)$ is  the $n-th$ Hermite polynomial and $A_n =
\frac {1} { \pi^{\frac {1} {4} } \; \sqrt{2^n n!}  }$ . The energy
is quantized according to:
\begin{equation}
\label{eq2.19}
E_{\perp,n} = e\, B \, (2 \, n + 1) \; .
\end{equation}
In order to solve Eq.~(\ref{eq2.13}), we expand $\omega(z)$ in
terms of the eigenstates of $\gamma^3$:
\begin{equation}
\label{eq2.20}
\omega(z) = \phi^1_+ \, u^1_+ + \phi^2_+ \, u^2_+ +
\phi^1_- \, u^1_- + \phi^2_- \, u^2_-
\end{equation}
where $u^s_{\pm}$ are given by
\begin{eqnarray}
\label{eq2.21}
 u_{\pm}^1 &=& \left(
\begin{array}{c}
1 \\ 0 \\ \pm i\\ 0
\end{array}
\right)\; , \\
\label{eq2.22}
 u_{\pm}^2 &=& \left(
\begin{array}{c}
0 \\ 1 \\ 0\\ \mp i
\end{array}
\right)\ \; ,
\end{eqnarray}
which satisfy  the following relations:
\begin{eqnarray}
\label{eq2.23}
\gamma^3 u_{\pm}^{1,2} &=& \pm i u_{\pm}^{1,2}
\; \; \; \; \; \; \; \;  \gamma^1\, u^1_{\pm} = \pm \, i \, u^2_{\mp}
\nonumber
\\
\gamma^0 u_{\pm}^{1,2}&=& + u_{\mp}^{1,2}
\; \; \; \; \; \; \; \; \;  \gamma^1\, u^2_{\pm} = \mp \, i \, u^1_{\mp}
\nonumber
\\
\gamma^1 \gamma^2 u_{\pm}^1 &=& -i u_{\pm}^1
\; \; \; \; \; \; \; \; \;  \gamma^2\, u^1_{\pm} = \mp  \, u^2_{\mp}
\\
\gamma^1 \gamma^2 u_{\pm}^2 &=& +i u_{\pm}^2 \, \; \; \;  \; \; \;
\; \; \;  \gamma^2\, u^2_{\pm} = \mp  \, u^1_{\mp} \: .
\nonumber
\end{eqnarray}
With the aid  of Eqs.(\ref{eq2.21})-(\ref{eq2.23}), we recast
Eq.~(\ref{eq2.13}) into:
\begin{equation}
\label{eq2.24}
\left( \frac {d^2} {d z^2} \mp \xi \frac {d g} {d
z} - \xi^2  g^2(z) + {E'}^2+e\, B \right) \phi_{\pm}^1=0 \; ,
\end{equation}
\begin{equation}
\label{eq2.25}
\left( \frac {d^2} {d z^2} \mp \xi \frac {d g} {d
z} - \xi^2 g^2(z) + {E'}^2 - e\, B\right) \phi_{\pm}^2=0 \; .
\end{equation}
For definiteness, let us consider Eq. (\ref{eq2.24}). Following
\cite{Ayala:1994mk}, we introduce the new variable
\begin{equation}
\label{eq2.26}
j= \frac {1} {1+ e^{2 \, z}} .
\end{equation}
It is easy to check that:
\begin{equation}
\label{eq2.27}
g(z)=1 - 2\,j \; ,
\end{equation}
\begin{equation}
\label{eq2.28}
\frac {d g} {d z} = 4 j \, (1-j) \; ,
\end{equation}
and
\begin{equation}
\label{eq2.29}
\frac {d^2} {d z^2}=4 \, j \, (1-j) \left[(1-2j) \frac {d} {d j} +
j \, (1-j) \, \frac {d^2} {d j^2} \right].
\end{equation}
Eqs.~(\ref{eq2.26})-(\ref{eq2.29}) allow us to rewrite
Eq.~(\ref{eq2.24}) as:
\begin{equation}
\label{eq2.30} \left[ \frac {d^2} {d j^2} + \frac {1-2\, j}
{j(1-j)} \frac {d} {d j} + \frac {{E'}^2 + eB\mp 4 \xi \,j (1-j) -
4 \, \xi^2 (1-2 \,j)^2} {4 j^2 (1-j)^2} \right] \phi^1_{\pm}= 0 \;
.
\end{equation}
By examining the behaviour of the differential equation near the
singular points $j=0$ and $j=1$, we find:
\begin{equation}
\label{eq2.31} \phi_{\pm}^1(z) \; = \; j^{\alpha^1} \; ( 1 \; - \;
j)^{\beta^1} \; \chi_{\pm}^1(j) \; ,
\end{equation}
where $\chi^1_{\pm}(j)$ is regular near the singularities. A
standard calculations gives:
\begin{equation}
\label{eq2.32} \alpha^1 =\frac{i}{2}\sqrt{{E'}^2 +eB-\xi^2}
\;\;\;{\phantom {aaaa}} \;\; , \;\; \beta^1 = \alpha^1 \; .
\end{equation}
Imposing that Eq.~(\ref{eq2.31}) is solution of
Eq.~(\ref{eq2.30}), we see that $\chi^1_{\pm}(j)$ satisfies
a hypergeometric equation~\cite{Gradshteyn:1963book} with
parameters given by:
\begin{eqnarray}
\label{eq2.33}
a_{\mp}^1 &=& 2 \, \alpha^1 + \frac {1} {2} -
\left|\xi \mp \frac {1} {2} \right|,
\\
\label{eq2.34}
b_{\mp}^1 &=& 2 \, \alpha^1 + \frac {1} {2} + \left|\xi \mp
\frac {1} {2} \right|,
\\
\label{eq2.35}
c^1 &=& 2 \alpha^1 +1 \; .
\end{eqnarray}
As well known, the general solution of the hypergeometric equation is given
by the combination of the two independent solutions:
\begin{eqnarray}
\label{eq2.36}
 _{2}F_{1}(a_{\mp},b_{\mp},c,j) \, ,
\\
\label{eq2.37}
j^{1-c} \; _{2}F_{1}(a_{\mp}+1-c,b_{\mp}+1-c,2-c;j) \; .
\end{eqnarray}
Therefore  Eq.~(\ref{eq2.31}) becomes:
\begin{eqnarray}
\label{eq2.38}
 \phi_{\pm}^1 &=& A^1_{\pm} \,  j^{\alpha^1} (1-j)^{\alpha^1} \,
 _2F_1(a_{\mp}^1,b_{\mp}^1,c^1,j) + \nonumber
\\
&& +  B_{\pm}^1 \, j^{-\alpha^1} (1-j)^{\alpha^1} \,
 _2F_1(a_{\mp}^1+1-c^1,b_{\mp}^1+1-c^1,2-c^1;j) \; ,
\end{eqnarray}
where $A_{\pm}^1$ and $B_{\pm}^1$ are normalization constants.
For simplicity we define:
\begin{eqnarray}
\label{eq2.39}
\phi_{\pm}^{(-\alpha^1)} &=& j^{\alpha^1}
(1-j)^{\alpha^1} \, _2F_1(a_{\mp}^1,b_{\mp}^1,c^1;j)\; ,
\\
\label{eq2.40}
\phi_{\pm}^{(+\alpha^1)} &=& j^{-\alpha^1} (1-j)^{\alpha^1} \,
  _2F_1(a_{\mp}^1+1-c^1,b_{\mp}^1+1-c^1,2-c^1;j) \; .
\end{eqnarray}
Then Eq.~(\ref{eq2.38}) can be written as:
\begin{equation}
\label{eq2.41}
\phi^1_{\pm} = A^1_{\pm} \, \phi_{\pm}^{(-\alpha^1)}
+ B^1_{\pm} \,  \phi_{\pm}^{(+\alpha^1)}.
\end{equation}
Therefore, assuming $\omega(z)= \phi^1_+ \, u^1_+$,  Eq.
(\ref{eq2.11}) becomes:
\begin{equation}
\label{eq2.42}
\chi(x,y,z)= A_n \, e^{- \frac {1} {2} \zeta^2 + i\, p_y \, y} \,
H_n(\zeta) \,  [A^1_+ \, \phi^{(- \alpha^1)}_++ B^1_+ \,
\alpha^{(+ \alpha^1)}_+] \, u^1_+.
\end{equation}
Finally, the  general solution the  can be obtained from
Eq.~(\ref{eq2.5}):
\begin{eqnarray}
\label{eq2.43}
\Psi_r^1(x,y,z,t) &=&
[(i \gamma^0 \, \partial_t+\xi g(z) + i \gamma^3
\partial_z)\, \phi^1_+ \, u_+^1 \, e^{- i E_r t}] \,  f(x,y) + \nonumber
\\
&  &
 [(i \gamma^1 \partial_x+ i \gamma^2 \partial_y - e \gamma^2 A_2) f(x,y)]
\, \phi^1_+ \, u_+^1\, e^{- i \, E_r \, t} \nonumber
\\
 & \equiv & \psi_r (z,t) \, f(x,y) + \eta(x,y) \, \phi^1_+ \, u^1_+ \,
e^{- i \,E_r \, t}
\end{eqnarray}
where
\begin{equation}
\label{eq2.44}
\psi_r (z,t) \equiv
[i \gamma^0 \, \partial_t+\xi g(z) + i \gamma^3
\partial_z]\, \phi^1_+ \, u_+^1 \, e^{- i E_r t} \; 
\end{equation}
and 
\begin{equation}
\label{eq2.45}
 \eta(x,y) \equiv
 (i \gamma^1 \partial_x+ i \gamma^2 \partial_y - e \gamma^2 A_2)
\, A_n \,H_n(\zeta) \,  e^{- \frac {1} {2} \zeta^2+ i \, p_y \, y}
\; .
\end{equation}
In order to esplicitate $\psi_r(z,t)$, let us assume that
\begin{equation}
\label{eq2.46} \varphi^1_{\pm} = \left[\mp \frac {d} {d z} + \xi
\, g(z) \right]\, \phi^1_{\pm} \; ,
\end{equation}
so that  Eq.~(\ref{eq2.44}) becomes:
\begin{equation}
\label{eq2.47} \psi_r(z, t)=(E_r \, \phi^1_+ \, u_-^1 \, +
\varphi^1_+\, u^1_+)\, e^{- i E_r t}.
\end{equation}
A calculation similar to the one performed in the Appendix of Ref.
\cite{Ayala:1994mk} gives:
\begin{equation}
\label{eq2.48}
\varphi^{(-\alpha^1)}_{\pm}=(\, \xi \, \pm 2 \, \alpha^1 \, ) \,
\phi^{(-\alpha^1)}_{\mp}
\end{equation}
and
\begin{equation}
\label{eq2.49} \varphi^{(+\alpha^1)}_{\pm}=(\, \xi \, \mp 2 \,
\alpha^1 ) \, \phi^{(+\alpha^1)}_{\mp} \; .
\end{equation}
We have then:
\begin{eqnarray}
\label{eq2.50}
\psi_r(z,t)&=& A^1 \, [E_r \, {\phi_+^1}^{(-\alpha^1)} \, u_-^1
            +(\xi+2 \, \alpha^1) {\phi_-^1}^{(-\alpha^1)} \, u_+^1]+
\nonumber
\\
& &            B^1 \, [E_r \, {\phi_+^1}^{(+\alpha^1)}
u_-^1
            +(\xi- 2 \, \alpha^1) {\phi_-^1}^{(+\alpha^1)} \, u_+^1]
\, e^{-i\, E_r\, t}.
\end{eqnarray}

We are now interested in the calculation of the reflection and
transmission coefficients for fermions incident on the wall from $z
\rightarrow - \infty$ to $z \rightarrow + \infty$. To this end, it
is necessary to consider the asymptotic forms of $\phi^{(-
\alpha^1)}_{\pm}$ and $\phi^{(+ \alpha^1)}_{\pm}$:
\begin{equation}
\label{eq2.51} \lim_{z \rightarrow +\infty} j^{\pm \alpha^1}
(1-j)^{\alpha^1} = \exp(\mp \, 2 \, \alpha^1 \, z)\, ,
\end{equation}
\begin{equation}
\label{eq2.52} \lim_{z \rightarrow -\infty} j^{\alpha^1}
(1-j)^{\pm \alpha^1} = \exp(\pm \, 2 \, \alpha^1 \, z)\, .
\end{equation}
Equation~(\ref{eq2.50}) has two terms with different asymptotic
properties for $z \rightarrow + \infty$ and $z \rightarrow -
\infty$. It is simple to see that the boundary conditions for
$\psi_r (z)$ imply  $A^1=0$. As a consequence we have:
\begin{equation}
\label{eq2.53} \psi_r(z,t)=B^1 \,[\, E_r \, {\phi_+^1}^{(+
\alpha^1)} \, u_-^1 + (\xi - 2 \, \alpha^1) \, {\phi_-^1}^{(+
\alpha^1)} \, u^1_+ ] e^{-i \, E_r\, t} \, .
\end{equation}
Concerning $\eta(x,y)$, since
\begin{equation}
\label{eq2.54} \frac {d} {d x} = \sqrt{eB} \frac {d} {d \zeta} \;
\; , \phantom{jaskdjs} \;\;\;\;\;\;\; \frac {d H_n(\zeta)} {d
\zeta} = 2 \, n \, H_{n-1} (\zeta)\; ,
\end{equation}
we get from  Eq.~(\ref{eq2.45})
\begin{equation}
\label{eq2.55} \eta(x,y) = A_n \, \{ i\,  \gamma^1 \, \sqrt{eB} \,
[2\, n \, H_n(\zeta) - \zeta \, H_n(\zeta)] - (p_y-eBx)\, \gamma^2
\, H_n(\zeta)\}\, e^{i p_y y - \frac {1} {2} \zeta^2} \; .
\end{equation}
Observing that $\omega(z) = \phi^1_+ \, u^1_+$ and taking into
account Eq.~(\ref{eq2.23}), we obtain
\begin{equation}
\label{eq2.56}
\eta(x,y) \, \phi^1_+\, u^1_+ \, e^{- i \, E_r \, t}
= A_n \, [( \zeta \, \sqrt{eB} + p_y - eBx)\, H_n(\zeta)
- 2 n \, \sqrt{eB} \, H_{n-1} (\zeta)] \, \phi^1_+ \, u^2_-
\, e^{ i p_y y - \frac {1} {2} \zeta^2 - i E_r t}.
\end{equation}
We can note now that the coefficient of $H_n$ in the last equation
vanishes. So that we get:
\begin{eqnarray}
\label{eq2.58}
\Psi_r(x,y,z,t)& =&  A_n \, B^1\,
\{H_n(\zeta) \, [ E_r \, \phi^{(+ \alpha^1)}_+ \, u^1_- +
(\xi - 2 \, \alpha^1) \, \phi_-^{(+ \alpha^1)} \, u^1_+] +  \nonumber
\\
& & - 2 n \sqrt{eB} \, H_{n-1} \, \phi_+^{(+\alpha^1)} \, u^2_-\}
\, e^{ i p_y y - \frac {1} {2} \zeta^2 - i E_r t} \; .
\end{eqnarray}
Writing
\begin{equation}
\label{eq2.59}
A_r (x,y,t) = A_n \,  B^1\,
 e^{ i p_y y - \frac {1} {2} \zeta^2 - i E_r t} \; ,
\end{equation}
the general solution becomes:
\begin{eqnarray}
\label{eq2.60}
\Psi_{r,n}^1(x,y,z,t) = A_r(x,y,t) \,
&\{&
[E_r\, H_n(\zeta) \, u_-^1 -
2 n \sqrt{eB} \, H_{n-1} (\zeta) \, u_-^2] \, \phi^{(+ \alpha^1)}_+
\nonumber
\\
& & + H_n(\zeta) \, (\xi - 2 \alpha^1) \, u^1_+ \, \phi^{(+
\alpha^1)}_- \, \}  \; .
\end{eqnarray}
To obtain the asymptotic behaviour of the wave function, we use Eq.
(\ref{eq2.39}) to get:
\begin{eqnarray}
\label{eq2.61}
\Psi_r(x,y,z,t) & =&  A_r(x,y,t)
\{ [E_r\, H_n(\zeta) \, u_-^1 -
2 n \sqrt{eB} \, H_{n-1} (\zeta) \, u_-^2] \;
 _{2}F_{1}(- \xi +1, \xi, 1-2\, \alpha^1; j) \, + \nonumber
\\
& &
+ H_n(\zeta) \, (\xi - 2 \alpha^1) \, u^1_+ \;
_{2}F_{1}(- \xi, \xi+1, 1-2\, \alpha^1; j)\}    
\, j^{-\alpha^1} (1-j)^{\alpha^1}
\, .
\end{eqnarray}
In order to find the expansion for $z \rightarrow - \infty$ (i.e.
$j \rightarrow 1$), we need to consider the analytical
extension of the hypergeometric functions by means of the  Kummer's
formula~\cite{Gradshteyn:1963book}:
\begin{eqnarray}
\label{eq2.62}
F(a,b,c;z) &=&  \frac {\Gamma (c) \, \Gamma (c-a-b)}
                    {\Gamma (c-a) \, \Gamma (c-b)}
                    \; F(a,b,a+b-c+1,z) + (1-z)^{c-a-b} \cdot \nonumber
                     \\ &&
                     \cdot \, \frac {\Gamma (c) \;
                    \Gamma (a+b-c)}
                    {\Gamma (a) \; \Gamma (b)} \;
                    F(c-a,c-b,c-a-b+1,1-z)\;.
\end{eqnarray}
By means of the well known relation:
\begin{equation}
\label{eq2.63} \Gamma(z+1)=z \; \Gamma(z) \; ,
\end{equation}
and after some manipulations we obtain the transmitted, incident
and reflected wave functions:
\begin{equation}
\label{eq2.64}
\left( \phantom{\frac {1} {2} } \Psi^1_{r,n} (x,y,z,t) \right)^{tran}
= A_r(x,y,t) [ \, H_n(\zeta) \, E_r\, u^1_- - 2n \sqrt{eB}
H_{n-1} (\zeta) \, u^2_- +
H_n (\zeta) \, (\xi - 2 \, \alpha^1) \, u_+^1 ] \, e^{2 \alpha^1\, z}
\end{equation}
\begin{eqnarray}
\label{eq2.65}
\left( \phantom{\frac {1} {2} } \Psi^1_{r,n} (x,y,z,t) \right)^{inc}
&  & =
A_r(x,y,t) \, \frac {\Gamma(1- 2 \, \alpha^1) \; \Gamma(- 2 \, \alpha^1)}
{\Gamma (- 2 \, \alpha^1 + \xi) \; \Gamma (- 2 \, \alpha^1 - \xi)}
\, e^{2 \, \alpha^1 \, z} \,
\left[ \frac {E_r \, H_n(\zeta)} {- 2 \, \alpha^1 - \xi}\, u^1_- \,\,  +\right.
\nonumber
\\
& &
+\left.
H_n(\zeta) \, u^1_+
- \frac {2 \, n \sqrt{eB} \, H_{n-1}(\zeta)} {- 2 \, \alpha^1 - \xi}\, u^2_-
\right]
\end{eqnarray}
\begin{eqnarray}
\label{eq2.66}
\left( \phantom{\frac {1} {2}}
\Psi^1_{r,n}(x,y,z,t)\right)^{refl} =
& &
A_r(x,y,t) \, \frac {\Gamma(1- 2 \, \alpha^1) \; \Gamma( 2 \, \alpha^1)}
{\Gamma (\xi) \; \Gamma (- \xi)}
\, e^{- 2 \, \alpha^1 \, z} 
\left[ \frac {E_r \, H_n(\zeta)} {- \xi}\, u^1_- \,\,   +\right. 
\nonumber
\\
& & 
+ \left. \frac
{H_n(\zeta) \, (\xi - 2 \, \alpha^1)} { \xi} \, u^1_+ 
- \frac {2 \, n \sqrt{eB} \, H_{n-1}(\zeta)} {- \xi}\, u^2_- \right] \; .
\end{eqnarray}
Making use of the following relations:
\begin{equation}
(u^1_-)^{\dag} \gamma^0 \gamma^3 u_-^1=0 \;\;\;\;\;\;\;\;
(u^1_+)^{\dag} \gamma^0 \gamma^3 u_-^1=-2 \, i \;\;\;\;\;\;\;\;\;
(u^2_-)^{\dag} \gamma^0 \gamma^3 u_-^1=0
\nonumber
\end{equation}
\begin{equation}
\label{eq2.67}
(u^1_-)^{\dag} \gamma^0 \gamma^3 u_+^1=2\, i
\;\;\;\;\;\;\;
(u^1_+)^{\dag} \gamma^0 \gamma^3 u_+^1=0
\;\;\;\;\;\;\;\;\;\;\;\;
(u^2_-)^{\dag} \gamma^0 \gamma^3 u_+^1=0
\end{equation}
\begin{equation}
(u^1_-)^{\dag} \gamma^0 \gamma^3 u_-^2=0 
\;\;\;\;\;\;\;\;\;\;
(u^1_+)^{\dag} \gamma^0 \gamma^3 u_-^2=0 
% \phantom {-2 \,i}
\;\;\;\;\;\;\;\;\;\;
(u^2_-)^{\dag} \gamma^0 \gamma^3 u_-^2=0 \; , \nonumber
\end{equation}
it is straightforward to obtain the currents:
\begin{equation}
\label{eq2.68}
(j^3_V)^{inc}= \frac {8}{\pi^2} i\, \alpha^1\,  E_r |A_r(x,y,t)|^2 \,
H_n^2(\zeta) \,
|\Gamma(1+ 2 \alpha^1) \Gamma(2 \alpha^1)|^2 \, \sin[\pi(2 \alpha^1+\xi)]
\sin[\pi(2 \alpha^1- \xi)]\, , 
\end{equation}
\begin{equation}
\label{eq2.69}
(j^3_V)^{tran}= - 8 \, i \, \alpha^1 \, |A_r(x,y,t)|^2 \,
H^2_n(\zeta) \, E_r \, ,
\end{equation}
\begin{equation}
\label{eq2.70}
(j^3_V)^{refl}=\frac {8}{\pi^2} i\, \alpha^1\,
E_r |A_r(x,y,t)|^2 \, H_n^2(\zeta) \, |\Gamma(1+ 2 \alpha^1)
\Gamma(2 \alpha^1)|^2 \, \sin^2(\pi \, \xi) \; .
\end{equation}
By means of the identity~\cite{Gradshteyn:1963book}
\begin{equation}
\label{eq2.71} \Gamma(z) \Gamma(-z) = \frac {- \pi} {z \sin (\pi
z)} \; ,
\end{equation}
we finally obtain the following reflection and transmission coefficients:
\begin{equation}
\label{eq2.71}
R^1 = \frac {- (j^3_V)^{refl}} {(j^3_V)^{inc}}= \frac {\sin^2(\pi
\, \xi)} {\sin[\pi(\xi- 2 \, \alpha^1)] \, \sin[\pi(\xi+2 \,
\alpha^1)]} \; ,
\end{equation}
\begin{equation}
\label{eq2.72}
T^1= \frac {(j^3_V)^{tran}} {(j^3_V)^{inc}}= \frac {- \sin^2(2 \pi
\, \alpha^1)} {\sin[\pi(\xi- 2 \, \alpha^1)] \, \sin[\pi(\xi+2 \,
\alpha^1)]} \; .
\end{equation}
It is worthwhile to stress that the dependence of the reflection 
and transmittion coefficients on the magnetic field is encoded into
$\alpha^1$. Indeed, from Eqs. (\ref{eq2.14}), (\ref{eq2.19}), and
(\ref{eq2.32}) we have:
\begin{equation}
\label{eq2.73}
\alpha^1 =\frac{i}{2}\sqrt{E^2 - 2neB -\xi^2}  \; \; \; \; \;
 n=0,1,2... \; \; .
\end{equation}
In the same way we can solve the case of fermions with spin
projection on the third spatial axis antiparallel to the magnetic
field. In this case the relevant equation turns out to be 
Eqs.~(\ref{eq2.24}) and (\ref{eq2.25}). 
Following the same steps of the previous derivation, we find:
\begin{equation}
\label{eq2.74}
R^2 = \frac {\sin^2(\pi \, \xi)} {\sin[\pi(\xi- 2 \, \alpha^2)] \,
\sin[\pi(\xi+2 \, \alpha^2)]} \; ,
\end{equation}
\begin{equation}
\label{eq2.75}
T^2 = \frac {- \sin^2(2 \pi \, \alpha^2)} {\sin[\pi(\xi- 2 \,
\alpha^2)] \, \sin[\pi(\xi+2 \, \alpha^2)]} \; ,
\end{equation}

where now:
\begin{equation}
\label{eq2.76}
\alpha^2 =\frac{i}{2}\sqrt{E^2 - 2eB(n+1) -\xi^2}   \; \; \; \; \;
 n=0,1,2... \; \; .
\end{equation}
It is interesting to note that the dependence due to the motion in
the direction transverse to the magnetic field plane  factorizes in the
expression of the currents, in such a way that the reflection and transmission
coefficients do not show any explicit $(x,y)$ dependence.
Of course, the reflection and transmission coefficients depend on
the Landau level index $n$. \\
In Fig.~1 we display the reflection coefficient for
parallel spin $R^{1}$  as a function of the scaled energy $
{E}/{\xi}$, with fixed magnetic field at various values of $n$. We
see that there is total reflection for fermions with parallel spin at
energies $E^2 -\xi^2 = 2neB$. In the case of antiparallel spin we
find that the total reflection occurs at energies $E^2 -\xi^2 =
2(n+1)eB$. As we shall discuss in the next Section, this peculiar
anomalous scattering can be understood as due to 
the presence of solutions localized on the kink domain wall. \\
Figure~2 shows both the reflection coefficients for parallel and
antiparallel spin as a function of the magnetic field  at fixed
energy and two different values of  Landau level index $n$. Note
that  $R^{1}$ for $n=0$ is independent of the magnetic field. We
see that the magnetic field is able to produce an asymmetry in the
spin distribution, but there is no particle-antiparticle
asymmetry, which would be relevant in the electroweak
baryogenesis. Moreover, we see that  the difference between these
coefficients grows with an increasing field strength. \\
Finally, in Fig.~3 we compare the transmission coefficients for
parallel and antiparallel spin as a function of the magnetic field
at fixed energy and two different values of  Landau level index
$n$. Note that, as expected, we have $R + T = 1$.
\renewcommand{\thesection}{\normalsize{\Roman{section}.}}
\section{\normalsize{SOLUTIONS LOCALIZED ON THE WALL}}
\renewcommand{\thesection}{\arabic{section}}
In this Section we discuss  fermion states corresponding to
solutions of the Dirac equation (\ref {eq2.4}) 
localized on the domain wall. Following the approach of the  previous
Section, we find again  Eqs.~(\ref {eq2.24}) and (\ref {eq2.25}), that for
completeness we rewrite here:
\begin{equation}
\label{eq3.1}
\left( \frac {d^2} {d z^2} \mp \xi \frac {d g(z)} {d z} - \xi^2 
g^2(z) + {E'}^2 + eB \right) \phi_{\pm}^1=0 \; ,
\end{equation}
\\
\begin{equation}
\label{eq3.2}
\left( \frac {d^2} {d z^2}\mp\xi\frac {d g(z)} {d z}-\xi^2 g^2(z)
+ {E'}^2 -eB \right) \phi_{\pm}^2=0 \; ,
\end{equation}
where, according to Eqs.~(\ref {eq2.14}) and (\ref {eq2.19}),
\begin{equation}
\label{eq3.3}
{E'}^2 \, = \, E^2 \, -  e\, B \, (2 \, n + 1) \; .
\end{equation}
It is known since long time that in the absence of magnetic field
these differential equations admit zero energy solutions localized
on the wall~\cite{Jackiw:1976fn,Niemi:1986vz}. In our case, we see
that the condition to localize  fermions on the wall is given by:
\begin{equation}
\label{eq3.4}
{E'}^2+eB=0 \;\;\;\;\; \text {for} \;\;\;\; \phi^1_{\pm}\, ,
\end{equation}
\begin{equation}
\label{eq3.5}
{E'}^2-eB=0 \;\;\;\; \; \; \; \text {for} \;\;\;\; \phi^2_{\pm} \; ,
\end{equation}
namely,
\begin{equation}
\label{eq3.6}
E^2 \, = \, 2 n e B \;\;\;\;\; \; \; \; \text {for} \;\;\;\;
\phi^1_{\pm} \, ,
\end{equation}
\begin{equation}
\label{eq3.7}
E^2 \, = \, 2 (n+1) e B \;\; \text {for} \;\;\;\; \phi^2_{\pm} \; .
\end{equation}
In Section~2 we found that there is total reflection for fermions
with parallel and antiparallel spin at energies $E^2 -\xi^2 =
2neB$ and $E^2 -\xi^2 = 2(n+1)eB$, respectively. The difference is
due to the fermion mass $\xi$ which, indeed, vanishes on the wall
where the system is in a symmetric phase. From a physical point of
view, we see that fermions with asymptotically  non-zero
momentum $|\vec{p}| $  equal to $\sqrt{2neB}$ for parallel spin
and $\sqrt{2(n+1)eB}$ for antiparallel spin can be trapped on the domain 
wall.
\\
Let us now discuss the localized solutions. Inserting
Eqs.~(\ref{eq3.4}) and~(\ref {eq3.5}) into Eqs.~(\ref{eq3.1}) and 
(\ref{eq3.2}), respectively, we get:
\begin{equation}
\label{eq3.8}
\left( \frac {d^2} {d z^2}\mp\xi\frac {d g(z)} {d z}-\xi^2
g^2(z)\right) \phi_{\pm}^{1,2}=0 \; .
\end{equation}
It is easy to find the solutions of Eq.~(\ref{eq3.8}):
\begin{equation}
\label{eq3.9}
\phi^{1,2}_{\pm}(z) = N \, \exp \left[\mp \xi \int_0^z d \, z' \,
g(z') \right] \, ,
\end{equation}
with $N$ normalization constant. Clearly, the $\phi^{1,2}_+$
solutions must be neglected because they are divergent for $|z|
\rightarrow +\infty$. We are left with the $\phi^{1,2}_-$
solutions, explicitly given by:
\begin{equation}
\label{eq3.10}
\phi_-^{1,2}(z) = N \,(\cosh z)^{- \xi}.
\end{equation}
In order to evaluate  the localized states, we insert Eq.~(\ref
{eq3.10}) into Eq.~(\ref {eq2.5}) and take care of Eqs.~(\ref
{eq2.7}),~(\ref {eq2.11}) and~(\ref {eq2.20}) to get:
\begin{equation}
\label{eq3.11}
\psi^{1,2}_{loc} \, (x,y,z,t)= (i \gamma^{\mu}
\partial_{\mu} + \xi \, g(z) - e \gamma^{\mu} A_{\mu}) \, f(x,y)
\, \phi^{1,2}_-(z) \, u^{1,2}_-\, e^{- i \, E t} \; .
\end{equation}
After some manipulations we obtain:
\begin{equation}
\label{eq3.12}
\psi^1_{loc} \, (x,y,z,t) = N \, A_n \, e^{- \frac {\zeta^2} {2} +
i \, p_y \, y
 - i \, E \, t} \, (\cosh z)^{-\xi} \,
[ H_n \, E \, u^1_+ +  2 \, n \, \sqrt{eB} \, H_{n-1} \, u^2_+ ]
\end{equation}
and
\begin{equation}
\label{eq3.13}
\psi^2_{loc} \, (x,y,z,t) = N \, A_{n} \, e^{- \frac {\zeta^2} {2}
+ i \, p_y \, y
 - i \, E \, t} \, (\cosh z)^{-\xi} \,
[ H_{n}\, E \, u^2_+ +  \sqrt{eB} \, H_{n+1} \, u^1_+ ] \; ,
\end{equation}
where $A_n$  has been defined in Sec~2. Of course,
$ E^2 \, = 2 \, n\, e \, B $  in Eq.~(\ref{eq3.12}), while
$ E^2 \, = 2 \, (n+1) \, e \,B $ in Eq.~(\ref{eq3.13}). \\
The normalization condition
\begin{equation}
\label{eq3.14}
\int d^3 x \; \psi^{\dag}_{loc} \; \psi_{loc} = \delta (p_y -
p'_y) \; \delta_{n n'}
\end{equation}
gives 
\begin{equation}
\label{eq3.15}
N^2= \frac {\sqrt{e B}} {8 \, \pi \, B(\xi,1/2) \, E^2} \, ,
\end{equation}
where $B(x,y)$ is the Bernoulli beta function~
\cite{Gradshteyn:1963book}. \\
It is interesting to note that our localized solutions can be
rewritten in the equivalent form:
\begin{eqnarray}
\label{eq3.16}
\psi_{loc} \, (x,y,z,t) &=& N \, \left(
\begin{array}{c}
v(x,y)\\ i \, \sigma^3\, v(x,y)
\end{array}
\right) \,  (\cosh z)^{-\xi} \, e^{-i \, E \, t} \; ,
\end{eqnarray}
where, for instance, in the case of the  localized wave function
$\psi^1_{loc}$ we find
\begin{eqnarray}
\label{eq3.17}
v(x,y)= A_n \, e^{i \, p_y \, y - \frac {\zeta^2} {2}}   \left(
\begin{array}{c}
H_n(\zeta) \\ \frac {2 \, n \, \sqrt{eB}} {E_r} \, H_{n-1}(\zeta)
\end{array}
\right) \; .
\end{eqnarray}
A similar result holds for $\psi^2_{loc}$.
\renewcommand{\thesection}{\normalsize{\Roman{section}.}}
\section{\normalsize{CONCLUSIONS}}
\renewcommand{\thesection}{\arabic{section}}
Much work has been recently devoted to study effects due to scattering of
particles from domain walls between the phases of broken and
unbroken symmetry at the electroweak phase
transition~\cite{Ayala:1994mk,Funakubo:1994en,Farrar:1994hn,Farrar:1995vp}.
The main effort of this paper consists in the investigation of the 
effects of a constant magnetic field on the scattering of fermions
on planar walls. In particular we focused our attention on kink domain walls,
which are of interest in a continuum cosmological phase transition.
We solved the Dirac equation for scattering of fermions off domain
walls,  and computed the transmission and reflection coefficients. As
expected, we find that the constant magnetic field induces a spin
asymmetry in fermion reflection and transmission. 
\\
The results
obtained, as a matter of fact, do not allow to produce directly
asymmetries in some local charges, which is a prerequisite for electroweak
baryogenesis~\cite{Trodden:1998ym,Riotto:1998bt,Riotto:1999yt}.
However, we would like to stress that, in the physical condition of
early Universe, fermions moving through the domain walls will
interact not only with the wall but also with the particles in the
surrounding plasma. So that, even though we expect in general that
an asymmetry between fermion and antifermion distributions in the
primordial plasma could be induced by magnetic fields, such
analysis requires the study of  quantum Boltzmann transport
equation~\cite{Kadanoff:1962book,Lifshitz:1979book}.
\\
An interesting aspect of kink domain walls, not shared by
bubble domain walls, is the presence of fermion states 
localized on the wall, effect due to the magnetic field. One could
speculate that for these trapped fermions the sphaleron mechanism
could efficiently converts any local charge asymmetry
into a baryon number asymmetry.  We deserve such an analysis to a future work. 
\\
Aside from these considerations, kink domains walls with the
associated magnetic field could display interesting cosmological
properties.
Let us then conclude this paper by briefly discussing the role of the
domain walls in the early Universe. As we have anticipated in the Introduction,
the possible role of
kink domain walls in the cosmological dynamics has been so far 
neglected, 
due to the fact that the existence of such walls was ruled out by
the Zel'dovich, Kobazarev and Okun~\cite{Zeldovich:1974uw}
argument. However, as 
we argued before, ferromagnetic domain walls with
vanishing effective surface tension~\cite{Cea:1998ep,Cea:1999jz}
are an open possibility. 
\\
It is interesting to stress that the same mechanism which leads
to a vanishing effective surface tension of the domain walls gives
rise to a vanishing effective energy-momentum tensor.
However, in Section~3 we have seen that kink domains in a constant
magnetic field display positive energy fermion  states localized
on the wall. If one takes into account the contribution of these
trapped fermions to the energy-momentum tensor, then one finds
that the kink domain wall acquires a non vanishing traceless
energy-momentum tensor. As a consequence, the gravitational
dynamics of such domain walls is fully relativistic. We shall
report progress on this subject  in a forthcoming
work~\cite{Campanelli:2001}.
%
%
%
%#############################################################
%#########       A  P  P  E  N  D  I  C  E      ##############
%#############################################################
\newpage
\appendix{\normalsize{\bf {APPENDIX}}}
\appendix\section{Bubble Walls}
In this Appendix we discuss fermions scattering off bubble walls in
a constant magnetic field. This case is relevant for a first order
phase transition where the conversion from one phase to another
takes place through nucleation. The region separating the two phases is
considered as the wall. Under the assumption that the wall is thin and
that the phase transition takes place when the energy densities of
both the phases are degenerate, it is possible to approximate the wall
profile in the form
\begin{equation}
\label{A1}
\varphi(z)=\frac {v} {2} [1+\tanh z] \: ,
\end{equation}
where we set the bubble wall thickness $\Delta=1$. We see that
$z < 0$ represents the region outside the bubble, i.e. the
region in the symmetric phase where particles are massless.
Conversely, for $z >  0$, the system is inside the bubble, i.e. 
in the broken phase and the particles have acquired a finite mass.
When scattering is not affected by diffusion, the problem of
fermion reflection and transmission through the wall can be casted
in terms of solving the Dirac equation with a position dependent
fermion mass, proportional to the Higgs field~\cite{Ayala:1994mk}.
\\
Recently, it has been shown that in the 
presence of primordial hypermagnetic fields it is possible to
generate an axial asymmetry during the reflection and transmission
of fermions off bubble walls \cite{Ayala:2001pj}. 
This result has been obtained with
the same approximation used in  our previous work~\cite{Cea:2000wz},
i.e. by considering solutions describing the motion od fermions
perpendicular to the wall. In this Appendix we include the effects
due to the motion of fermions parallel to the wall in the presence of
a constant magnetic field. However, our results can be easily
extended to the case of hypermagnetic fields
discussed in  Ref.~\cite{Ayala:2001pj}. \\
Starting from the Dirac equation (\ref{eq2.2}) and assuming
Eq.~(\ref{eq2.6}), we obtain:
\begin{eqnarray}
\label{A2}
& & [ 
- \partial^2 + i \, \xi \gamma^3 \partial_z \, g(z) - 2 i \, e
\,A^{\mu}  \partial_{\mu} - i \, e
\partial^{\mu} \, A_{\mu}- \frac {e} {2} \,
\sigma^{\mu \nu} \, F_{\mu \nu}  +
\nonumber
\\
& &  -  \xi^2 \, g^2(z) + e^2 \, A_{\mu}
A^{\mu} ] \, \Phi(x,y,z,t) = 0 \; ,
\end{eqnarray}
where now:
\begin{equation}
\label{A3}
g(z)=\frac {v} {2} [1+\tanh z] \; ,
\end{equation}
and $2 \xi$ is the mass that the fermion acquires in the broken
phase. \\
Using Eq.~(\ref{eq2.10}) and factorizing $\Phi(x,y,z,t)$ according
to Eqs.~(\ref{eq2.7}) and~(\ref{eq2.11}), we get:
\begin{equation}
\label{A4}
[ \, \partial_x^2 + \partial_y^2 - 2 \, i \, e \, A^2
\,
\partial_y - e^2 \, (A_2)^2 + E_{\perp} \, ] \, f(x,y)= 0
\end{equation}
and
\begin{equation}
\label{A5}
[\, \partial^2_z + i \, \xi \, \gamma^3 \,
\partial_z \, g(z) - \xi^2 \, g^2(z) + {E'}^2 + i \, e \, B \,
\gamma^1 \, \gamma^2 \, ] \, \omega(z) =0
\end{equation}
with
\begin{equation}
\label{A6} E^2={E'}^2\, + E_{\perp} \; .
\end{equation}
Eq.~(\ref{A4}) agrees with Eq.~(\ref{eq2.12}). So that
$f(x,y)$ is given by Eqs.~(\ref{eq2.14}) and (\ref{eq2.18}), and
$E_{\perp}$
is quantized according to Eq.~(\ref{eq2.19}). \\
In order to solve Eq.~(\ref{A4}), we expand $\omega(z)$ in terms of the
spinors $u_{\pm}^s$  eigenstates of $\gamma^3$,
Eqs.~(\ref{eq2.21}) and (\ref{eq2.22}). We get:
\begin{equation}
\label{A7}
 \left( \frac {d^2} {d z^2}\mp\xi\frac {d g} {d z} -
\xi^2  g^2(z) + {E'}^2 + eB \right) \phi_{\pm}^1=0
\end{equation}
\\
\begin{equation}
\label{A8}
\left( \frac {d^2} {d z^2}\mp\xi\frac {d g} {d z}- \xi^2 g^2(z) +
{E'}^2 -eB \right) \phi_{\pm}^2=0 \; .
\end{equation}
For definiteness, let us consider Eq.~(\ref{A7}). Introducing the
new variable
\begin{equation}
\label{A8bis}
j= \frac {1} {1+ e^{2 z}} \; ,
\end{equation}
we obtain
\begin{equation}
\label{A9}
\left[ \frac {d^2} {d j^2} + \frac {1-2j} {j(1-j)} \frac {d} {d j}
+ \frac {{E'}^2 + eB \mp 4 \xi j(1-j) -4 \xi^2(1-j)^2} {4 j^2
(1-j)^2}  \right] \phi_{\pm}^1=0 \; .
\end{equation}
The behaviour of the differential equation near the singular
points allows us to write
\begin{equation}
\label{A10}
\phi_{\pm}^1(z) \; = \; j^{\alpha^1} \; ( 1 \; - \; j)^{\beta^1}
\;
 \chi_{\pm}^1(j) \; ,
\end{equation}
where
\begin{equation}
\label{A11} \alpha^1=\frac {i} {2} \sqrt{E^2 -2neB - 4 \, \xi^2}\, ,
\end{equation}
\begin{equation}
\label{A12}
\beta^1=\frac {i} {2} \,\sqrt{E^2 -2neB} \; .
\end{equation}
It is easy to see that $\chi_{\pm}^1(j)$ satisfy the
hypergeometric equation~\cite{Gradshteyn:1963book} with
parameters:
\begin{eqnarray}
\label{A13}
 a_{\mp}^1 &=& \alpha^1+\beta^1 + \frac {1} {2} -
\left|\xi \mp \frac {1} {2} \right|, \nonumber
\\
b_{\mp}^1 &=& \alpha^1 + \beta^1 + \frac {1} {2} + \left|\xi \mp
\frac {1} {2} \right|, \nonumber
\\
c^1 &=& 2 \alpha^1 +1 \; .
\end{eqnarray}
Concerning the solution of Eq.~(\ref{A8}), it is straightforward
to see that $\phi_{\pm}^2$ can be obtained from $\phi_{\pm}^1$
provided $\alpha^1$ and $\beta^1$ are replaced by
\begin{equation}
\label{A14} \alpha^2=\frac {i} {2} \sqrt{E^2 -2(n+1)eB - 4 \,
\xi^2}\, ,
\end{equation}
\begin{equation}
\label{A15}
\beta^2=\frac {i} {2} \,\sqrt{E^2 -2(n+1)eB} \; .
\end{equation}
In order to write the most general solutions, it is useful to
introduce the following notations:
\begin{equation}
\label{A16}
{\epsilon_s} =
             \begin{cases}
           + 1\;\; \text{if}  \;\; s=1
              \\
           - 1\;\; \text{if}  \;\; s=2
             \end{cases}  ,
\;\;\;\;
 h_{s} =
             \begin{cases}
           2n \frac {H_{n-1}} {H_n} \;\; \text{if}  \;\; s=1
              \\
            \frac {H_{n+1}} {H_n} \;\; \text{if}  \;\; s=2
             \end{cases} , 
\;\;\;\;
 \bar{s} =
             \begin{cases}
           + 1 \;\; \text{if}  \;\; s=2
              \\
           - 1 \;\; \text{if}  \;\; s=1 \; .
             \end{cases}
\end{equation}
After some calculations similar to those performed in  Section~2,  
we obtain the
transmitted, incident and reflected wave functions:
\begin{equation}
\label{A17}
(\Psi^{\pm}_{n,r,s})^{tran}= A_r \, H_n[E_r \, u^s_{\mp}+2(\xi\mp
\alpha^s)u^s_{\pm} \mp \sqrt{eB} \, h_s \, u^{\bar{s}}_{\mp}] \,
\exp(2 \, \alpha^s \, z) \; ,
\end{equation}
\begin{eqnarray}
\label{A18}
(\Psi^{\pm}_{n,r,s})^{inc} & = & A_r\, H_n \, \frac {\Gamma(1-2 \,
\alpha^s)  \Gamma(-2 \beta^s)} {\Gamma(- \alpha^s- \beta^s+ \xi)
\, \Gamma(- \alpha^s - \beta^s - \xi)} \frac {1} {- \alpha^s -
\beta^s \mp \xi} \nonumber \\ & & \left[ E_r u^s_{\mp} + \frac
{2(\xi \mp \alpha^s)(- \alpha^s- \beta^s \mp \xi)} {- \alpha^s -
\beta^s \pm \xi} \, u^s_{\pm} \mp \sqrt{eB} \, h_s \,
u^{\bar{s}}_{\mp} \right] e^{2 \beta^s z} \; ,
\end{eqnarray}
\begin{eqnarray}
\label{A19}
(\Psi^{\pm}_{n,r,s})^{refl} & = & A_r\, H_n \, \frac {\Gamma(1-2 \,
\alpha^s)  \Gamma(2 \beta^s)} {\Gamma(- \alpha^s+ \beta^s- \xi) \,
\Gamma(- \alpha^s + \beta^s + \xi)} \frac {1} {- \alpha^s +
\beta^s \mp \xi} \nonumber \\ & & \left[ E_r u^s_{\mp} + \frac
{2(\xi \mp \alpha^s)(- \alpha^s+ \beta^s \mp \xi)} {- \alpha^s +
\beta^s \pm \xi} \, u^s_{\pm} \mp \sqrt{eB} \, h_s \,
u^{\bar{s}}_{\mp} \right] e^{- 2 \beta^s z} \; .
\end{eqnarray}
Likewise, the transmitted, incident and reflected currents are:
\begin{equation}
\label{A20}
(j^3_{V,s})^{tran}= 8 \, \epsilon_s \, E_r \, H^2_n \, |A_r|^2 \,
|\alpha^s|
\end{equation}
\begin{equation}
\label{A21}
(j^3_{V,s})^{inc} = \frac {- 8} {\pi^2} \epsilon_s E_r H^2_n |A_r|^2
\, |\alpha^s| |\Gamma(1+2 \alpha^s) \Gamma(2 \beta^s)|^2 \,
\sin[\pi(\xi+\alpha^s+\beta^s)] \, \sin[\pi(\xi - \alpha^s-
\beta^s)]
\end{equation}
\begin{equation}
\label{A22}
 (j^3_{V,s})^{refl} = \frac {8} {\pi^2} \epsilon_s E_r H^2_n
|A_r|^2 \, |\alpha^s| |\Gamma(1+2 \alpha^s) \Gamma(2 \beta^s)|^2 \,
\sin[\pi(\xi+\alpha^s-\beta^s)] \, \sin[\pi(\xi - \alpha^s+
\beta^s)] \; .
\end{equation}
Therefore the reflection and transmission coefficients are given by:
\begin{equation}
\label{A23}
R^s= \frac {\sin[\pi(\alpha^s - \beta^s + \xi)] \,
\sin[\pi(-\alpha^s + \beta^s + \xi)]} {\sin[\pi(-\alpha^s -
\beta^s + \xi)] \, \sin[\pi(+\alpha^s + \beta^s + \xi)]}\, ,
\end{equation}
\begin{equation}
\label{A24}
T^s= \frac {- \sin(2\pi\, \alpha^s) \,  \sin(2 \pi\beta^s)}
{\sin[\pi(-\alpha^s - \beta^s+ \xi)] \, \sin[\pi(+\alpha^s +
\beta^s + \xi)]} \; .
\end{equation}
%
%
%
%
%++++++++++++++++++++++++++++++++++++++++++++++++++++++++++++++++
%
%
%\bibliography{wall}

%
%
\vfill
\newpage
\begin{figure}[H]
\label{Fig1}
\begin{center}
\includegraphics[clip,width=0.85\textwidth]{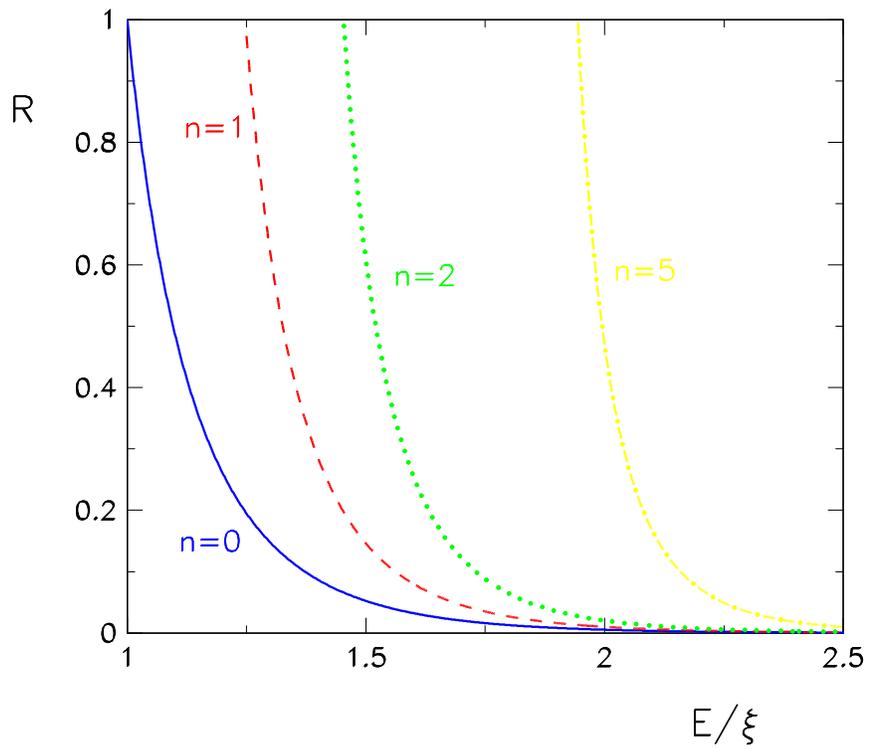}
%{figure_01.eps}
\caption{   
Reflection coefficient for parallel spin  versus the
scaled energy for four different values of the Landau level index
$n$ with $e B = 0.1$ . }
\end{center}
\end{figure}

\begin{figure}[H]
\label{Fig1}
\begin{center}
\includegraphics[clip,width=0.85\textwidth]{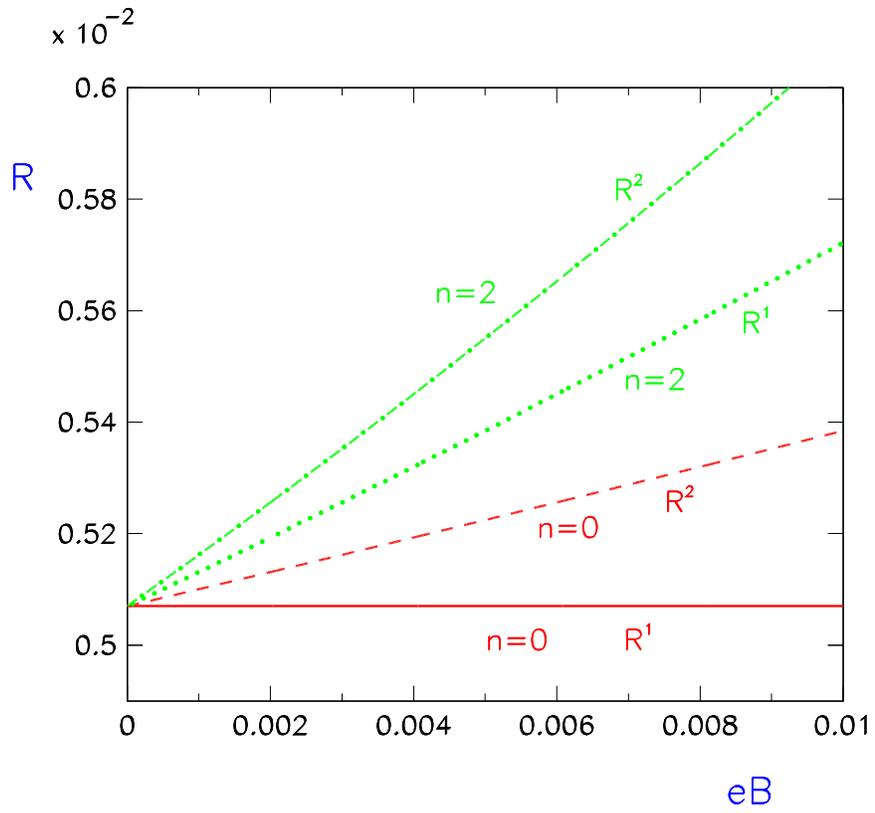}
%{figure_02.eps}
\caption{Reflection coefficients for parallel and antiparallel
spin at fixed energy as a function of $eB$ for two different
values of $n$.}
\end{center}
\end{figure}

\begin{figure}[H]
\label{Fig1}
\begin{center}
\includegraphics[clip,width=0.85\textwidth]{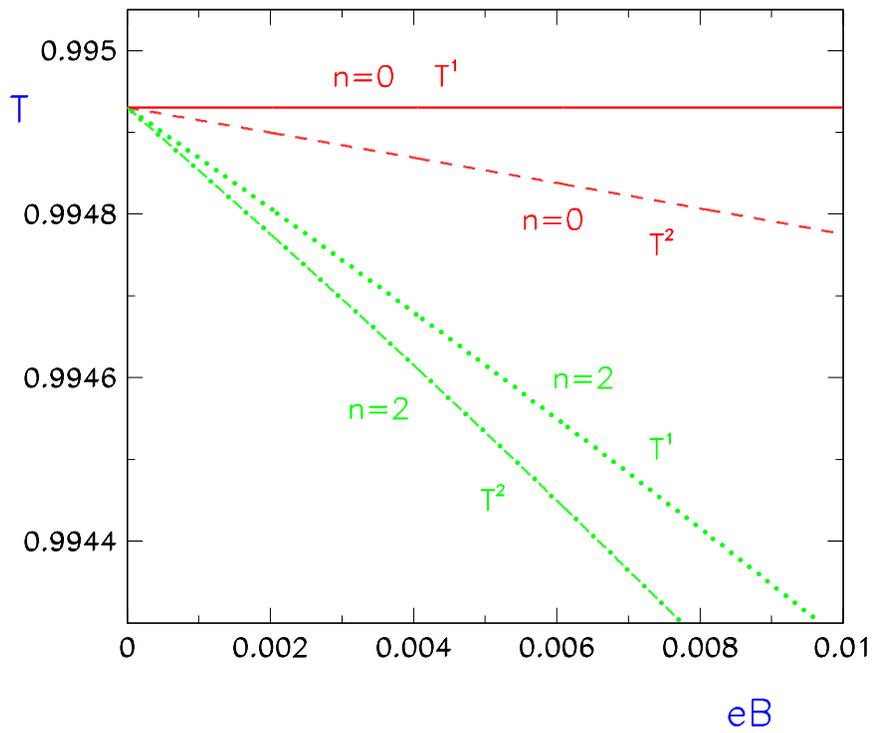}
\caption{Transmission coefficients for parallel and antiparallel
spin at fixed energy as a function of $eB$ for two different
values of $n$.}
\end{center}
\end{figure}

\end{document}